\documentclass[12pt, onecolumn]{IEEEtran}
\usepackage{amsmath}
\usepackage{amssymb}
\usepackage{cite}
\usepackage{epsfig}
\usepackage{dsfont}

\renewcommand{\baselinestretch}{1.4}

\newtheorem{thm}{Theorem}
\newtheorem{lem}[thm]{Lemma}

\newtheorem{col}{Corollary}

\newcommand{\tr}{{\rm trace}}
\newcommand{\diag}{{\rm diag}}

\begin{document}
%
%
%
%
%
%
%
%
\title{Matrix-Lifting Semi-Definite Programming for Decoding in Multiple Antenna Systems}
\author{Amin Mobasher and Amir K. Khandani
\thanks{The authors are with Coding \& Signal Transmission Laboratory
(www.cst.uwaterloo.ca), Department of Electrical and Computer
Engineering, University of Waterloo, Waterloo, Ontario, Canada, N2L
3G1, e-mail: \{amin, khandani\}@cst.uwaterloo.ca}
\thanks{This work is partly presented in the $10^{th}$ Canadian
Workshop on Information Theory (CWIT'07), Edmonton, AB, June 2007.}
}
\maketitle

\begin{abstract}
This paper presents a computationally efficient decoder for multiple
antenna systems. The proposed algorithm can be used for any
constellation (QAM or PSK) and any labeling method. The decoder is
based on matrix-lifting Semi-Definite Programming (SDP). The
strength of the proposed method lies in a new relaxation algorithm
applied to the method of \cite{MoTaSoKh05}. This results in a
reduction of the number of variables from $(NK+1)^2$ to $(2N+K)^2$,
where $N$ is the number of antennas and $K$ is the number of
constellation points in each real dimension. Since the computational
complexity of solving SDP is a polynomial function of the number of
variables, we have a significant complexity reduction. Moreover, the
proposed method offers a better performance as compared to the best
quasi-maximum likelihood decoding methods reported in the
literature.
\end{abstract}

\section{Introduction}

The problem of Maximum Likelihood (ML) decoding in Multi-Input
Multi-Output (MIMO) wireless systems is known to be NP-hard. A
variety of sub-optimum polynomial time algorithms based on
Semi-Definite Programming (SDP) are suggested for MIMO
decoding~\cite{StLuWo03, StLuWo031, LuLuKi03, MaChDi04, WiElSh05L,
SidLuo06, YaZhZhXu07, MoTaSoKh05, MoTaSoKh05C3}. The first quasi ML
decoding methods based on SDP were introduced for PSK
signalling~\cite{StLuWo03, StLuWo031, LuLuKi03, MaChDi04}, offering
a near ML performance and a polynomial time worst case complexity.
Subsequently, SDP methods were used for decoding of MIMO systems
based on QAM constellation~\cite{WiElSh05L,MoTaSoKh05}.

The method presented in \cite{WiElSh05L} is for MIMO systems using
16-QAM, where the structure of constellation is captured by a
polynomial constraint. Then, by introducing some slack variables,
the constraints are expressed in terms of quadratic polynomials.
This method can be generalized for larger constellations at the cost
of defining more slack variables, increasing the complexity, and
significantly decreasing the performance. The method proposed in
\cite{SidLuo06} is a further relaxation of \cite{WiElSh05L}, only
utilizing upper and lower bounds on the symbol energy in the
relaxation step. There is a very slight degradation in performance
compared to \cite{WiElSh05L}; however, its computational complexity
is independent of the constellation size for any uniform QAM (order
of complexity is cubic). The method in \cite{YaZhZhXu07} is a
further tightening of \cite{SidLuo06} by appending some inequality
conditions that are implicit in the alphabet constraint. Its
computational complexity is still less than \cite{WiElSh05L}.

In \cite{MoTaSoKh05}, an efficient approximate ML decoder for MIMO
systems is developed based on vector lifting SDP. The transmitted
vector is expanded as a linear combination (with zero-one
coefficients) of all the possible constellation points in each
dimension. Using this formulation, the distance minimization in
Euclidean space is expressed in terms of a binary quadratic
minimization problem. The minimization of this problem is over the
set of all binary rank-one matrices with column sums equal to one.
Although the algorithm in \cite{MoTaSoKh05} is a sub-optimal
decoding method, it is shown that by adding several extra
constraints, it can approach the ML performance. However,
implementing the extra constraints increases the computational
complexity.

In this paper, we introduce a new algorithm based on matrix-lifting
SDP \cite{DinWol06, Bek07} for any constellation (QAM or PSK) and
any labeling method. This algorithm is inspired by the method in
\cite{MoTaSoKh05} with an efficient implementation resulting in a
better performance and lower computational complexity. In SDP
optimization problems, the computational complexity is a polynomial
function of the number of variables. Using the proposed method, the
number of variables in \cite{MoTaSoKh05} is decreased from
$(NK+1)^2$ to $(2N+K)^2$, where $N$ is the number of antennas and
$K$ is the number of constellation points in each real dimension. In
addition to this large reduction in the complexity, simulation
results show that the proposed algorithm also outperforms all other
known convex quasi-ML decoding methods, e.g. \cite{WiElSh05L,
SidLuo06, YaZhZhXu07}.

Following notations are used in the sequel. The space of $ N \times
K $ (resp.~$ N \times N $) real matrices is denoted by ${\mathcal
M}_{N \times K}$ (resp.~${\mathcal M}_{N}$), and the space of $ N
\times N $ symmetric matrices is denoted by $ {\mathcal S}_N $. For
a $N\times K$ matrix $ \mathbf{X} \in {\mathcal M}_{N \times K} $,
the $(i,j)$th element is represented by $x_{ij}$, where $1\leq i
\leq N,\; 1\leq j \leq K$, i.e. ${\bf X}=[x_{ij}]$. We use
${\tr}(\mathbf{A})$ to denote the trace of a square matrix ${\bf
A}$. The space of symmetric matrices is considered with the trace
inner product $\langle \mathbf{A},\mathbf{B} \rangle =
{\tr}(\mathbf{AB})$. For $ \mathbf{A}, \mathbf{B} \in {\mathcal
S}_{N}$, $ \mathbf{A} \succeq 0$ (resp.~$ \mathbf{A} \succ 0$)
denotes positive semi-definiteness (resp. positive definiteness),
and $ \mathbf{A} \succeq \mathbf{B}$ denotes  $
\mathbf{A}-\mathbf{B} \succeq 0$. For two matrices $ \mathbf{A},
\mathbf{B} \in {\mathcal M}_{N}$, $ \mathbf{A} \geq \mathbf{B} $, ($
\mathbf{A} > \mathbf{B} $) means $a_{ij}\geq b_{ij}$, $(a_{ij}
> b_{ij})$ for all $i,j$. The Kronecker product of two matrices $
\mathbf{A} $ and $ \mathbf{B} $ is denoted by $ \mathbf{A} \otimes
\mathbf{B} $. For $ \mathbf{X} \in {\mathcal M}_{N \times K}$, ${\rm
vec}(\mathbf{X})$ denotes the vector in $\mathbb{R}^{NK}$ (real
$NK$-dimensional space) that is formed from the columns of the
matrix $ \mathbf{X} $. For $\mathbf{X}\in {\mathcal M}_{N}$, ${\rm
diag}(\mathbf{X})$ is a vector of the diagonal elements of
$\mathbf{X}$. We use $\mathbf{e}_N \in \mathbb{R}^N$ (resp. $
\mathbf{0}_N  \in \mathbb{R}^N $) to denote the $ N \times 1 $
vector of all ones (resp. all zeros), $ \mathbf{E}_{N \times K} \in
{\mathcal M}_{N \times K}$ to denote the matrix of all ones, and $
\mathbf{I}_N $ to denote the $N\times N$ Identity matrix. For $
\mathbf{X} \in {\mathcal M}_{N \times K}$, the notation $
\mathbf{X}(1:i,1:j) $, $i<K$ and $j<N$ denotes the sub-matrix of $
\mathbf{X}$ containing the first $i$ rows and the first $j$ columns.

The rest of the paper is organized as follows. The problem
formulation is introduced in Section II. Section III is the review
of the vector-lifting semi-definite programming presented in
\cite{MoTaSoKh05}. In Section IV, we propose our new algorithm based
on matrix-lifting semi-definite programming. We use the geometry of
the relaxation to find a projected relaxation which has a better
performance. In Section V, we present an optimization method, based
on matrix nearness to find the integer solution of the original
decoding problem from the relaxed optimization problem. Finally,
Section VI conclude the paper with some simulation results.

\section{Problem Formulation}

A MIMO system with $\tilde{M}$ transmit antennas and $\tilde{N}$
receive antennas can be modeled by
\begin{equation}\label{eq:realchan}
{\bf y}={\bf H}{\bf x}+{\bf n},
\end{equation}
where $M=2\tilde{M}$, $N=2\tilde{N}$, ${\bf y}$ is the $M\times 1$
received vector, ${\bf H}$ is $M \times N$ real channel matrix,
${\bf n}$ is $N\times 1$ additive white Gaussian noise vector, and
${\bf x}$ is $N \times 1$ data vector whose components are selected
from the set $\{s_1, \cdots ,s_K\}$, see \cite{MoTaSoKh05}.
Noting $ x_i \in \{s_1, \cdots ,s_K\} $, for $ i=1,\cdots,N $, we
have
\begin{equation}\label{eq:idear}
x_i=u_{i,1}s_1 + u_{i,2}s_2 + \cdots + u_{i,K}s_K,
\end{equation}
where
\begin{equation}\label{eq:idcond}
u_{i,j} \in \left\lbrace 0,1 \right\rbrace \;\; {\rm and } \;\;
\sum_{j=1}^K u_{i,j}=1, \;\; \forall\, i=1,\cdots,N.
\end{equation}
Let
\begin{eqnarray}
\nonumber {\bf U} = \left[
\begin{array}{ccccccc}
u_{1,1} & \cdots & u_{1,K}\\
u_{2,1} & \cdots & u_{2,K}\\
\vdots & \ddots & \vdots\\
u_{N,1} & \cdots & u_{N,K}
\end{array}
\right] & \textmd{and} & {\bf s}=\left[
\begin{array}{c}
s_1\\
\vdots\\
s_K
\end{array}
\right].
\end{eqnarray}
Therefore, the transmitted vector is $ \mathbf{x} = \mathbf{U}
\mathbf{s}$ where ${\bf U}{\bf e}_K = {\bf e}_N$.

At the receiver, the ML decoding is given by
\begin{equation}\label{eq:harddec}
\hat{\mathbf{x}}=\arg \min_{x_i\in \{s_1, \cdots ,s_K\}} \|
\hat{\mathbf{y}} - \mathbf{H} \mathbf{x} \|^2,
\end{equation}
where $ \hat{\mathbf{x}} $ is the most likely input vector and
$\hat{\mathbf{y}} $ is the received vector. Noting ${\bf x}={\bf
Us}$, this problem is equivalent to
\begin{eqnarray}
\nonumber &&\min_{\mathbf{Ue}_K=\mathbf{e}_N} \|
\hat{\mathbf{y}}-\mathbf{HUs} \|^2 \equiv\\
&&\min_{\mathbf{Ue}_K=\mathbf{e}_N} \mathbf{s}^T \mathbf{U}^T
\mathbf{H}^T \mathbf{HUs} - 2\hat{\mathbf{y}}^T \mathbf{HUs}.
\end{eqnarray}
Therefore, the decoding problem can be formulated as
\begin{eqnarray}\label{eq:dec}
\nonumber \min && \mathbf{s}^T \mathbf{U}^T
\mathbf{H}^T \mathbf{HUs} - 2\hat{\mathbf{y}}^T \mathbf{HUs}\\
\nonumber s.t. && {\bf Ue}_K={\bf e}_N\\
&& u_{i,j} \in \{0,1\}.
\end{eqnarray}

Let $ \mathbf{Q} =\mathbf{H}^T \mathbf{H} $, ${\bf S}={\bf ss}^T$, $
\mathbf{C} = -\mathbf{s\hat{y}}^T \mathbf{H} $, and let ${\mathcal
E}_{N \times K}$ denote the set of all binary matrices in ${\mathcal
M_{N\times K}}$ with row sums equal to one, i.e.
\begin{equation}\label{eq:decfes}
{\mathcal E}_{N \times K} \hspace{-2pt} = \hspace{-2pt} \left\lbrace
{\bf U} \hspace{-2pt} \in \hspace{-2pt} {\mathcal M_{N\times K}}:
{\bf U}{\bf e}_K = {\bf e}_N, u_{ij} \in \{0,1\}\right\rbrace.
\end{equation}
Therefore, the minimization problem (\ref{eq:dec}) is
\begin{eqnarray}\label{eq:decsdp}
\nonumber \min && \tr \left({\bf SU}^T{\bf QU} + 2{\bf CU} \right)\\
s.t. && {\bf U} \in {\mathcal E}_{N \times K}
\end{eqnarray}

\section{Vector-Lifting Semi-Definite Programming}

In order to solve the optimization problem \eqref{eq:decsdp}, the
authors in \cite{MoTaSoKh05} proposed a \emph{quadratic vector
optimization} solution by defining ${\bf u}={\rm vec}({\bf U}^T),
{\bf U} \in {\mathcal E}_{N \times K}$. By using this notation, the
objective function is replaced by ${\bf u}^T({\bf Q} \otimes {\bf
S}) {\bf u} + 2{\rm vec}({\bf C})^T{\bf u}$. Then,
the quadratic form is linearized using the vector
\begin{math}
\left[
\begin{array}{c}
1\\
{\bf u}
\end{array} \right],
\end{math}
i.e.
\begin{eqnarray}
\nonumber {\bf Z}_{\bf u} &=& \left [
\begin{array}{c}
1 \\ {\bf u}
\end{array}
\right ]\left [
\begin{array}{cc}
1 & {\bf u}^T
\end{array}
\right ]\\ &=& \left [
\begin{array}{cc}
1 & {\bf u}^T \\ {\bf u} & {\bf u} {\bf u}^T
\end{array}
\right ] = \left [
\begin{array}{cc}
1 & {\bf u}^T \\ {\bf u} & {\bf X}
\end{array}
\right ],
\end{eqnarray}
where ${\bf X}={\bf u} {\bf u}^T$ and it is relaxed to ${\bf X}
\succeq {\bf u} {\bf u}^T$, or equivalently, by the Schur
complement, to the lifted constraint
\begin{math}
\left [
\begin{array}{cc}
1 & {\bf u}^T \\ {\bf u} & {\bf X}
\end{array}
\right ] \succeq 0.
\end{math}
Note that this matrix is selected from the set
\begin{equation}
{\mathcal F} := {\rm conv}\left \{ {\bf Z}_{\bf u}: {\bf u} = {\rm
vec}({\bf U}^T), ~{\bf U}\in {\mathcal E}_{N\times K} \right \},
\end{equation}
where ${\rm conv(.)}$ denotes the convex hull of a set. Therefore,
the decoding problem using \emph{vector lifting semi-definite
programming} can be represented by
\begin{eqnarray}\label{eq:vecsdp}
\nonumber \tr && \left [
\begin{array}{cc}
0 & {\rm vec}({\bf C})^T \\ {\rm vec}({\bf C}) & \mathbf{Q} \otimes
\mathbf{S}
\end{array}
\right ]\left [
\begin{array}{cc}
1 & {\bf u}^T \\ {\bf u} & {\bf X}
\end{array}
\right ]\\
s.t. && \left [
\begin{array}{cc}
1 & {\bf u}^T \\ {\bf u} & {\bf X}
\end{array}
\right ] \in \mathcal{F},
\end{eqnarray}
which can be solved by SDP technique.

Note that in \eqref{eq:vecsdp}, the optimization parameter is a
matrix in $\mathcal{S}_{NK+1}$, which has $(NK+1)^2$ variables. In
the following, we reduce the number of optimization variables by
exploiting the matrix structure of ${\bf U}$.

\section{Matrix-Lifting Semi-Definite Programming}

To keep the matrix ${\bf U}$ in its original form in
\eqref{eq:decsdp}, the idea is to use the constraint ${\bf X}={\bf
U}^T{\bf U}$ instead of ${\bf X}={\bf u} {\bf u}^T$. As a result, the
relaxation ${\bf X} \succeq {\bf u} {\bf u}^T$ is transformed to
${\bf X} \succeq {\bf U}^T{\bf U}$, or equivalently, by the Schur
complement,
\begin{math}
\left [
\begin{array}{cc}
{\bf I}_N & {\bf U} \\ {\bf U}^T & {\bf X}
\end{array}
\right ] \succeq 0.
\end{math}
This is known as matrix-lifting semi-definite programming.
Define the new variable ${\bf V}={\bf US}$. Since the matrix ${\bf
S}$ is symmetric, the objective function in \eqref{eq:decsdp} can be
represented as the Quadratic Matrix Program \cite{Bek07}
\begin{eqnarray}\label{eq:msdp}
\nonumber &\tr& \left( \left[
\begin{array}{cc}
{\bf U}^T & {\bf V}^T
\end{array}
\right] \left[
\begin{array}{cc}
{\bf 0} & \frac{1}{2}{\bf Q}\\
\frac{1}{2}{\bf Q} & {\bf 0}
\end{array}\right]\left[
\begin{array}{c}
{\bf U}\\
{\bf V}
\end{array}
\right] + 2{\bf CU} \right)\\
\nonumber = &\tr& \left( \left[
\begin{array}{cc}
{\bf 0} & \frac{1}{2}{\bf Q}\\
\frac{1}{2}{\bf Q} & {\bf 0}
\end{array}\right]\left[
\begin{array}{c}
{\bf U}\\
{\bf V}
\end{array}
\right] \left[
\begin{array}{cc}
{\bf U}^T & {\bf V}^T
\end{array}
\right] + 2{\bf CU} \right)\\
= &\tr& \left(\mathcal{L}_{\bf Q} {\bf W}_{\bf U}\right),
\end{eqnarray}
where
\begin{equation}
\mathcal{L} = \left[
\begin{array}{ccc}
{\bf 0} & {\bf C} & {\bf 0}\\
{\bf C}^T & {\bf 0} & \frac{1}{2}{\bf Q}\\
{\bf 0} & \frac{1}{2}{\bf Q} & {\bf 0}
\end{array}
\right]
\end{equation}
and
\begin{equation}\label{eq:const}
{\bf W}_{\bf U} = \left[
\begin{array}{ccc}
{\bf I} & {\bf U}^T & {\bf V}^T\\
{\bf U} & {\bf UU}^T & {\bf UV}^T\\
{\bf V} & {\bf VU}^T & {\bf VV}^T
\end{array}
\right].
\end{equation}
To linearize ${\bf W}_{\bf U}$, we consider the matrix
\begin{equation}\label{eq:optquad}
\left[
\begin{array}{c}
{\bf U}\\
{\bf V}
\end{array}
\right] \left[
\begin{array}{cc}
{\bf U}^T & {\bf V}^T
\end{array}
\right] = \left[
\begin{array}{cc}
{\bf X} & {\bf Y}\\
{\bf Y} & {\bf Z}
\end{array}
\right],
\end{equation}
where ${\bf X,Y,Z} \in \mathcal{S}_N$. This equality can be relaxed
to
\begin{equation}
\left[
\begin{array}{cc}
{\bf UU}^T & {\bf UV}^T\\
{\bf VU}^T & {\bf VV}^T
\end{array}
\right] - \left[
\begin{array}{cc}
{\bf X} & {\bf Y}\\
{\bf Y} & {\bf Z}
\end{array}
\right] \preceq 0.
\end{equation}
It can be shown that this relaxation is convex in the L\"{o}wner
partial order and it is equivalent to the linear constraint
\cite{DinWol06}
\begin{equation}
{\bf W}\triangleq\left[
\begin{array}{ccc}
{\bf I} & {\bf U}^T & {\bf V}^T\\
{\bf U} & {\bf X} & {\bf Y}\\
{\bf V} & {\bf Y} & {\bf Z}
\end{array}
\right] \succeq 0.
\end{equation}
On the other hand, the feasible set in \eqref{eq:decsdp} is the set
of binary matrices in $\mathcal{M}_{N\times K}$ with row sum equal
to one, the set $\mathcal{E}_{N\times K}$ in \eqref{eq:decfes}. By
relaxing the rank-one constraint for the matrix variable in
\eqref{eq:msdp}, we have a tractable SDP problem. The feasible set
for the objective function in \eqref{eq:msdp} is approximated by
\begin{eqnarray}\label{eq:msdpfes}
\nonumber \mathcal{F}_{\mathcal{M}} = {\rm conv} \left\{ {\bf
W}_{\bf U} \right. \hspace{-5pt} &|& \hspace{-5pt} {\bf U}
\in {\mathcal M_{N\times K}}: {\bf U}{\bf e}_K = {\bf e}_N,\\
\hspace{-5pt} && \hspace{-5pt} \left. u_{ij} \in \{0,1\}, \forall
i,j; {\bf V}={\bf US} \right\}
\end{eqnarray}
Therefore, the decoding problem can be represented by
\begin{eqnarray}\label{eq:MSDPimp}
\nonumber \min && \tr \left( \mathcal{L} {\bf W} \right)\\
s.t. && {\bf W} \in \mathcal{F}_{\mathcal{M}}.
\end{eqnarray}
Note that the size of matrix ${\bf W}$ is $(2N+K) \times (2N+K)$,
compared to $(NK+1) \times (NK+1)$ in \cite{MoTaSoKh05}. In SDP
optimization problems, the computational complexity is a polynomial
function of the number of variables (elements of ${\bf W}$). By the
new implementation of \eqref{eq:MSDPimp}, the number of variables in
\cite{MoTaSoKh05} is decreased from $(NK+1)^2$ to $(2N+K)^2$,
resulting in a large reduction in the complexity.

Although the rank constraint in \eqref{eq:optquad} is relaxed, we
can still consider some additional linear constraints to further improve
the quality of the solution. These constraints are valid for the
non-convex rank-constrained decoding problem. However, we force the
SDP problem to satisfy these constraints. Consider the auxiliary
matrix ${\bf V}$ and the symmetric matrices ${\bf X, Y}$ and ${\bf
Z}$ in matrix ${\bf W}$. Since ${\bf U} \in \mathcal{E}_{N\times K}$
and
\begin{math}
\sum_{j=1}^N u_{ij}^2=1,
\end{math}
it is clear that $\diag(X)={\bf e}_N$. Also, ${\bf Y}$ represents
${\bf USU}^T$ and ${\bf Z}$ represents ${\bf US^2U}^T$. It is easy
to show that

\begin{eqnarray}\label{eq:quadterm}
\diag(\bf Y) = {\bf U} \diag(\bf S) & {\rm and} & \diag(\bf Z) =
{\bf U} \diag(\bf S^2).
\end{eqnarray}
Moreover, ${\bf S}={\bf ss}^T$ (rank-one matrix) and $ {\bf
S}^2=(\sum_{1=i}^K s_i^2){\bf S}$. Therefore, instead of $\diag(\bf
Z) = {\bf U} \diag(\bf S^2)$, we have a stronger result for ${\bf
Z}$, i.e. $ {\bf Z}=(\sum_{1=i}^K s_i^2){\bf Y} $. Therefore, we
have
\begin{eqnarray}\label{eq:qapsdp}
\nonumber \min && \tr \left( \mathcal{L} \left[
\begin{array}{ccc}
{\bf I} & {\bf U}^T & {\bf V}^T\\
{\bf U} & {\bf X} & {\bf Y}\\
{\bf V} & {\bf Y} & {\bf Z}
\end{array}
\right] \right)\\
\nonumber s.t. && {\bf Ue}_K={\bf e}_N \;\;;\;\; {\bf U} \geq 0\\
\nonumber && {\bf V}={\bf US}\\
\nonumber && \diag({\bf X})={\bf e}_N\\
\nonumber &&\diag(\bf Y) = {\bf U} \diag(\bf S)\\
\nonumber && {\bf Z}=(\sum_{1=i}^K s_i^2){\bf Y}\\
\nonumber && \left[
\begin{array}{ccc}
{\bf I} & {\bf U}^T & {\bf V}^T\\
{\bf U} & {\bf X} & {\bf Y}\\
{\bf V} & {\bf Y} & {\bf Z}
\end{array}
\right] \succeq 0\\
&& {\bf U, V} \in \mathcal{M}_{N \times K}, {\bf X,Y, Z}\in
\mathcal{S}_N
\end{eqnarray}
The equation in \eqref{eq:quadterm} determines the diagonal elements
of ${\bf Y}$. This property is hidden in the special structure of
${\bf U}$, i.e. ${\bf U} \in \mathcal{E}_{N \times K}$. By using
this property, we can even add more constraints. The equation ${\bf
Y}={\bf USU}^T$ implies that $Y_{ij}=S_{kl}$ for some $k$ and $l$.
Therefore, the value of $Y_{ij}$ is between the minimum and the
maximum elements of ${\bf S}$. In addition, it can be easily shown
that in communication applications, ${\bf S}$, ${\bf Y}$, and ${\bf
Z}$ are diagonal dominant matrices (since ${\bf s}^T{\bf e}_K=0$).
This property can be also used to add more constraints to improve
the quality of the solution. Our studies show that the improvements
due to including the above constraints are marginal. Therefore, in
the sequel, we focus on the form given in \eqref{eq:qapsdp} with the
following consideration. The objective function in \eqref{eq:decsdp}
is $\tr \left({\bf SU}^T{\bf QU} + 2{\bf CU} \right)$ which is
equivalent to $\tr \left({\bf QU}{\bf SU}^T + 2{\bf UC} \right)$.
Exchanging the role of ${\bf Q}$ and ${\bf S}$ results in two
different formulations. Here, the auxiliary variable ${\bf V}$ is
defined as ${\bf QU}$. Similarly, the auxiliary variables ${\bf X,
Y,}$ and ${\bf Z}$ represents ${\bf U}^T{\bf U}, {\bf U}^T{\bf QU},$
and ${\bf U}^T{\bf Q}^2{\bf U}$, respectively. Therefore, it is easy
to show that the equivalent minimization problem is

\begin{eqnarray}\label{eq:equlsdp}
\nonumber \min && \tr \left( \left[
\begin{array}{ccc}
{\bf 0} & {\bf C} & {\bf 0}\\
{\bf C}^T & {\bf 0} & \frac{1}{2}{\bf S}\\
{\bf 0} & \frac{1}{2}{\bf S} & {\bf 0}
\end{array}
\right] \left[
\begin{array}{ccc}
{\bf I} & {\bf U} & {\bf V}\\
{\bf U}^T & {\bf X} & {\bf Y}\\
{\bf V}^T & {\bf Y} & {\bf Z}
\end{array}
\right] \right)\\
\nonumber s.t. && {\bf Ue}_K={\bf e}_N \;\;;\;\; {\bf U} \geq 0\\
\nonumber && {\bf V}={\bf QU}\\
\nonumber && \diag({\bf X})={\bf U}^T{\bf e}_N\;\; ; \;\; X_{ij}=0
\; i\neq j \\
\nonumber && {\bf Ye}_K = {\bf U}^T {\bf Qe}_N\;\; ; \;\; \tr({\bf
YE}_K)=\tr({\bf QE}_N)\\
\nonumber && {\bf Ze}_K = {\bf U}^T {\bf Q}^2{\bf e}_N\;\; ; \;\;
\tr({\bf ZE}_K)=\tr({\bf Q}^2{\bf E}_N)\\
\nonumber && \left[
\begin{array}{ccc}
{\bf I} & {\bf U} & {\bf V}\\
{\bf U}^T & {\bf X} & {\bf Y}\\
{\bf V}^T & {\bf Y} & {\bf Z}
\end{array}
\right] \succeq 0\\
&& {\bf U, V} \in \mathcal{M}_{N \times K}, {\bf X,Y, Z}\in
\mathcal{S}^K,
\end{eqnarray}
where the size of the variable matrix is $(2K+N)$. Note that both
\eqref{eq:qapsdp} and \eqref{eq:equlsdp} are equivalent, however,
depending on the structure of the system (values of $N$ and $K$), we
can use the one which offers a smaller number of variables. In the
following, we focus on \eqref{eq:qapsdp}, which is a better choice
for $N \leq K$.

\subsection{Geometry of the Relaxation}

In this section, we eliminate the constraints defining ${\bf
Ue}_K={\bf e}_N$ by providing a tractable representation of the
linear manifold spanned by this constraint. This method is called
\emph{gradient projection} or \emph{reduced gradient method}
\cite{HaReWo92}. The following lemma is on the representation of
matrices having sum of the elements in each row equal to one. This
lemma is used in our reduced gradient method.
\begin{lem}~\cite{MoTaSoKh05} \label{lem:row} Let
\begin{math}
{\bf G} = \left [
\begin{array}{c|c}
{\bf I}_{K-1} & -{\bf e}_{K-1}
\end{array} \right ] \in {\mathcal M}_{(K-1) \times K}
\end{math}
and
\begin{math}
{\bf F} = \frac{1}{K} \left( {\bf E}_{N\times K} - {\bf E}_{N\times
(K-1)} {\bf G} \right)  \in {\mathcal M}_{N \times K}.
\end{math}
A matrix ${\bf U} \in \mathcal{M}_{N \times K}$ with the property
that the summation of its elements in each row is equal to one, i.e.
$ {\bf Ue}_K = {\bf e}_N$, can be written as
\begin{equation}\label{eq:Ureduc}
{\bf U} = {\bf F} + \bf{\hat U} {\bf G},
\end{equation}
where $ \bf{\hat U} = {\bf U}(1:N,1:(K-1))$.
\end{lem}
\begin{col}
$\forall {\bf U} \in {\mathcal E}_{N\times K}$, $ \exists \hat{\bf
U}\in {\mathcal M}_{N\times (K-1)}$, $\hat{u}_{ij}\in \{0,1\}$ s.t.
${\bf U}={\bf F}+\bf{\hat U}{\bf G}$, where $\bf{\hat U}={\bf U}
(1:N,1:(K-1))$. Note that the summation of each row of $\bf{\hat U}$
is 0 or 1.
\end{col}

Consider the minimization problem \eqref{eq:decsdp}. By substituting
\eqref{eq:Ureduc}, the objective function is
\begin{eqnarray}
\nonumber && \tr \left({\bf SU}^T{\bf QU} + 2{\bf CU} \right)\\
\nonumber &=& \tr \left( {\bf S(F+\hat{U}G)}^T{\bf Q(F+\hat{U}G)} +
2{\bf C(F+\hat{U}G)} \right)\\
\nonumber &=& \tr \left({\bf GSG}^T \bf{\hat U}^T {\bf Q}\bf{\hat U}
+ {\bf GSF}^T {\bf Q} \bf{\hat U} + {\bf QFSG}^T \bf{\hat U}^T \right.\\
\nonumber && \qquad\quad \left. +{\bf GC} \bf{\hat U} + {\bf C}^T
{\bf G}^T \bf{\hat U}^T + 2{\bf CF} + {\bf SF}^T{\bf QF} \right)\\
&=& \tr \left( \hat{\mathcal L} \bf{W}_{\bf{\hat U}} + 2{\bf CF} +
{\bf SF}^T{\bf QF} \right),
\end{eqnarray}
where
\begin{eqnarray}
\nonumber \hat{\mathcal L} \hspace{-3pt}&=&\hspace{-3pt} \left[
\begin{array}{ccc}
{\bf 0} & {\bf GSF}^T {\bf Q}+{\bf GC} & {\bf 0}\\
{\bf QFSG}^T+{\bf C}^T{\bf G}^T & {\bf 0} & \frac{1}{2}{\bf Q}\\
{\bf 0} & \frac{1}{2}{\bf Q} & {\bf 0}
\end{array} \hspace{-3pt}
\right],\\
\nonumber {\bf W}_{\bf{\hat U}} \hspace{-3pt}&=&\hspace{-3pt} \left[
\begin{array}{ccc}
{\bf I} & \bf{\hat U}^T & \bf{\hat V}^T\\
\bf{\hat U} & \bf{\hat U}\bf{\hat U}^T & \bf{\hat U}\bf{\hat V}^T\\
\bf{\hat V} & \bf{\hat V}\bf{\hat U}^T & \bf{\hat V}\bf{\hat V}^T
\end{array}
\right],\\
\bf{\hat V} \hspace{-3pt}&=&\hspace{-3pt}  \bf{\hat U} {\bf GSG}^T.
\end{eqnarray}
Therefore, \eqref{eq:dec} can be written as
\begin{eqnarray}\label{eq:sdpred}
\nonumber \min && \tr  \left( \hat{\mathcal L} \bf{W}_{\bf{\hat U}}
\right)\\
\nonumber s.t. && \bf{\hat U}={\bf U} (1:N,1:(K-1)); \;\; {\bf U}
\in {\mathcal E}_{N \times K}\\
&& \bf{\hat V} = \bf{\hat U} \left({\bf GSG}^T  \right)
\end{eqnarray}
Using a similar procedure, we can show that
\eqref{eq:sdpred} is equivalent to the following reduced
matrix-lifting semi-definite programming problem:

\begin{eqnarray}\label{eq:reducsdp}
\nonumber \min && \tr \left( \hat{\mathcal L} \left[
\begin{array}{ccc}
{\bf I} & \bf{\hat U}^T & \bf{\hat V}^T\\
\bf{\hat U} &  \bf{\hat X} &  \bf{\hat Y}\\
\bf{\hat V} &  \bf{\hat Y} &  \bf{\hat Z}
\end{array}
\right] \right)\\
\nonumber s.t. &&  \bf{\hat U}{\bf e}_{K-1} \leq {\bf e}_N \;\;;\;\;
\bf{\hat U} \geq 0\\
\nonumber && \bf{\hat V}=\bf{\hat U} \left({\bf GSG}^T\right) \\
\nonumber && \diag(\bf{\hat X})= \bf{\hat U}{\bf e}_{K-1}\\
\nonumber && \diag(\bf{\hat Y})= \bf{\hat U} \diag\left({\bf
GSG}^T\right)\\
\nonumber && \bf{\hat Z}=\left(\sum_{1=i}^{K-1}
(s_i-s_K)^2\right)\bf{\hat Y}\\
\nonumber && \left[
\begin{array}{ccc}
{\bf I} & \bf{\hat U}^T & \bf{\hat V}^T\\
\bf{\hat U} &  \bf{\hat X} &  \bf{\hat Y}\\
\bf{\hat V} &  \bf{\hat Y} &  \bf{\hat Z}
\end{array}
\right] \succeq 0\\
&& \bf{\hat U},\bf{\hat V} \in \mathcal{M}_{N \times (K-1)},
\bf{\hat X},\bf{\hat Y}, \bf{\hat Z} \in \mathcal{S}_N
\end{eqnarray}
Note that this method can also be applied to the equivalent
formulation in \eqref{eq:equlsdp}.

\subsection{Solving the SDP Problem}

The relaxed decoding problems can be solved using Interior-Point
Methods (IPMs), which are the most common methods for solving SDP
problems of moderate sizes with polynomial computational
complexities \cite{AlHaOv98}. There are a large number of IPM-based
solvers to handle SDP problems, e.g., DSDP \cite{BenYe04}, SeDuMi
\cite{Stu01}, SDPA \cite{KoFuNaYa02}, etc. In our numerical
experiments, we use SDPA solver.

In the matrix-lifting SDP optimization problem \eqref{eq:qapsdp},
the rank-constrained matrix ${\bf W}_{\bf u}$ is relaxed to the
positive semi-definite matrix ${\bf W}$. Utilizing the
rank-constrained property of the variable parameter, the relaxed
problem \eqref{eq:qapsdp} can be solved using a non-linear
method, known as the \emph{augmented Lagrangian algorithm}. This
approach can be used for large problem sizes and the
complexity can be significantly reduced, while the
performance degradation is negligible~\cite{MobKha071}.

\section{Integer Solution - Matrix Nearness Problem}

Solving the relaxed decoding problems results in the solution
$\bf{\tilde U}$. In general, this matrix is not in
$\mathcal{E}_{N\times K}$. The condition ${\bf U}{\bf e}_K={\bf
e}_N$ is satisfied. However, the elements are between 0 and 1. This
matrix has to be converted to a 0-1 matrix by finding a matrix in
$\mathcal{E}_{N\times K}$ which is nearest to this matrix.
Matrix approximation problems typically measure the
distance between matrices with a norm. The Frobenius and spectral
norms are common choices as they are analytically
tractable.

To find the nearest solution in $\mathcal{E}_{N \times K}$
to $\bf{\tilde U}$, the solution of the relaxed problem, we solve
\begin{equation}
\min_{{\bf U} \in \; \mathcal{E}_{N\times K}} \|{\bf U} - \bf{\tilde
U}\|_{\mathds F}^2,
\end{equation}
where $\|{\bf A}\|_{\mathds F}^2$ is the Frobenius norm of the
matrix ${\bf A}$ which is defined as $\|{\bf A}\|_{\mathds F}^2 =
\tr ({\bf AA}^T)$, and

\begin{eqnarray}
\nonumber \|{\bf U} - \bf{\tilde U}\|_{\mathds F}^2 &=& \tr \left(
({\bf U} - \bf{\tilde U})( {\bf U} - \bf{\tilde U})^T \right)\\
&=& N - 2\tr(\bf{\tilde U}{\bf U}^T) + \tr(\bf{\tilde U}\bf{\tilde
U}^T).
\end{eqnarray}
The last equality is due to the fact that for any ${\bf U} \in
\mathcal{E}_{N\times K}$, we have $\diag({\bf UU}^T)={\bf e}_N$, see
\eqref{eq:qapsdp}. Therefore, after removing the constants, finding
the integer solution is the solution of the following problem:
\begin{equation}\label{eq:intsol}
\max_{{\bf U} \in \; \mathcal{E}_{N\times K}} \tr(\bf{\tilde U}{\bf
U}^T)
\end{equation}
Consider the maximization problem
\begin{eqnarray}\label{eq:realsol}
\nonumber \max && \tr(\bf{\tilde U}{\bf U}^T)\\
\nonumber s.t. && {\bf U}{\bf e}_K={\bf e}_N\\
&& 0 \leq {\bf U} \leq 1,
\end{eqnarray}
where $\leq$ in the last constraint is element-wise. This problem is
a linear programming problem with linear constraints and the optimum
solution is a corner point meaning that the constraints are satisfied
with equality at the optimum point. In other words, at the optimum
point, ${\bf U} \in \mathcal{E}_{N\times K}$. Therefore, to
find the solution for \eqref{eq:intsol}, we can simply solve the
linear problem \eqref{eq:realsol}, which is strongly polynomial
time. To improve this result, the randomization
algorithms, introduced in \cite{MoTaSoKh05}, can be further applied.

\section{Simulation Results}

\begin{figure}[htbp]
\centering
\includegraphics[scale=0.65]{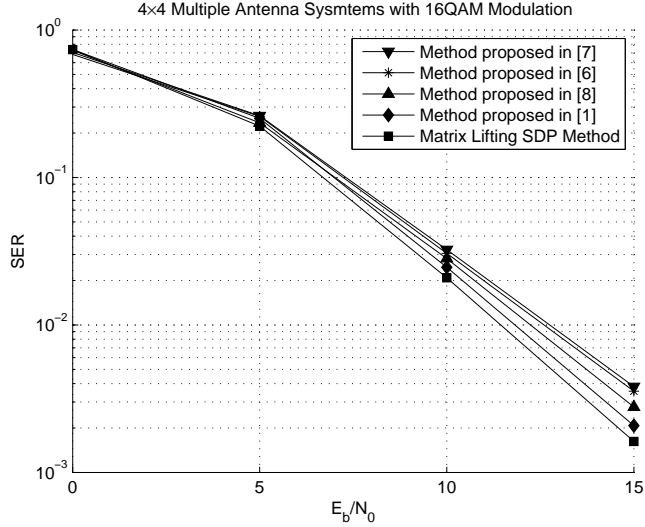}\\
\caption{Performance of the proposed matrix lifting SDP method in a
MIMO system with $4$ transmit and $4$ receive antennas employing
16-QAM} \label{fig:comp}
\end{figure}

We simulate the proposed matrix lifting method \eqref{eq:reducsdp}
for system with $4$ transmit and $4$ receive
antennas employing 16-QAM. Fig. \ref{fig:comp} shows
the performance of the proposed method vs. the performance of the
vector lifting method in \cite{MoTaSoKh05} and the previous known
methods in \cite{WiElSh05L, SidLuo06, YaZhZhXu07}. As it can be
seen, the proposed method outperforms all other convex sub-optimal
methods.

The worst case complexity of the proposed method solved by IPMs is a
polynomial function of the number of antennas (similar to the
analysis in \cite{MoTaSoKh05}). In the optimization problem of
\eqref{eq:qapsdp}, where $N\leq K$, the dimension of the matrix
variable ${\bf W}$ is $m=O(K)$ and the number of constraints is
$p=O(K^2)$. Similar to \cite{MoTaSoKh05}, it can be easily seen that
a solution to \eqref{eq:reducsdp} can be found in at most
$O(K^{5.5})$ arithmetic operations (utilizing the sparsity of the
rank-one constraint matrices), where the computational complexity of
\cite{MoTaSoKh05}, \cite{WiElSh05L}, \cite{YaZhZhXu07},
\cite{SidLuo06} are $\mathcal{O}(N^{5.5}K^{5.5})$,
$\mathcal{O}(N^{6.5}K^{6.5})$, $\mathcal{O}(K^2N^{4.5}+K^3N^{3.5})$,
and $\mathcal{O}(N^{3.5})$ respectively\footnote{Due to space limit,
we refer the reader to \cite{MoTaSoKh05} for a comparison on the
execution time of different methods.}. Note that for the equivalent
optimization problem \eqref{eq:equlsdp}, where $K \leq N$, the
computational complexity is at most $\mathcal{O}(N^{5.5})$. It must
be emphasized that depending on values of $N$ and $K$, we can
implement the optimization problem \eqref{eq:qapsdp} or
\eqref{eq:equlsdp} which results in less computational complexity.

Note that many of the constraints have very simple
structures. This property can be used to develop an interior-point
optimization algorithm fully exploiting the constraint structures of
the problem, thereby getting complexity order
better than that of using a general purpose solver such as SeDuMi or
SDPA. Moreover, we can further reduce the complexity of the proposed
method by implementing the augmented Lagrangian method
\cite{MobKha071}.

%
%
\section*{Acknowledgement}

The authors would like to thank R. Sotirov, H. Wolkowicz and Y. Ding
for many helpful discussions and comments. The authors would also
like to thank T. Davidson, M. Kisialiou, W. Ma and N. Sidiropoulos
for their comments on an earlier draft of this work.

\end{document}